\documentclass{llncs}
\usepackage{amscd}
\usepackage{graphicx}
\begin{document}
\title{Design Issues of JPQ: a Pattern-based Query Language for Document Databases}
%\thanks{This research is partially supported by the Fundamental Research Funds for the
%Central Universities of China under contract No.6082010, the Wuhan
%ChenGuang Youth Sci.\&Tech. Project under contract No.200850731369, the 111 Project under contract No.B07037, and the NSF of China under contract No.60688201.}}
\author{Xuhui Li\inst{1},  Mengchi Liu\inst{1}, Xiaoying Wu\inst{1} and Jinguang Gu\inst{2}}% and Shanfeng Zhu \inst{3}}
\authorrunning{Li, Liu, et-al.}
\institute{Computer School, Wuhan Univ.,\\
Wuhan, China\\ \email{lixuhui@whu.edu.cn, charleslzq@gmail.com, mengchi@sklse.org, xiaoying.wu@gmail.com} \and
School of Computer Sci. and Tech., Wuhan Univ. of Sci. and Tech., \\Wuhan, China\\
\email{simon@wust.edu.cn} %\and
%School of Computer Science and Shanghai Key Lab of Intelligent Information Processing, Fudan Univ. \\Shanghai, China\\
%\email{zhusf@fudan.edu.cn}
} \maketitle

\begin{abstract}
Document databases are becoming popular, but how to present complex document query  to obtain useful information from the document remains an important topic to study.
In this paper, we describe the design issues of a pattern-based document database query language named JPQ. JPQ uses various expressive patterns to extract and construct document fragments following a JSON-like document data model. It adopts tree-like extraction patterns with a coherent pattern composition mechanism to extract data elements from hierarchically structured documents and maintain the logical relationships among the elements. Based on these relationships, JPQ deploys a deductive mechanism to declaratively specify the data transformation requests and considers also data filtering on hierarchical data structure. We use various examples to show the features of the language and to demonstrate its expressiveness and declarativeness in presenting complex document queries.

\end{abstract}
\section{Introduction}
Document databases are a kind of so-called NoSQL databases which use nested collections of records or rows to store information. The data in the documents are organized in an \emph{aggregate}-oriented way where an aggregate is a collection of related objects. An aggregate is treated as a unit of manipulation meanwhile it  has allowable structure and types to enable flexible accessing. Document databases often organize large amounts of aggregates by associating them with the keys to be accessed efficiently and semantically. The dominant document format is JSON\cite{json} which is widely used as a lightweight data-interchange language.

A JSON-like document database is quite different from conventional key-value stores where the aggregates are usually opaque blobs of meaningless bits to be parsed and processed by applications. It is also different from XML database, even though they both exhibit a hierarchical style and they both use semantic labels, i.e., tags or keys, to associate contents, because the keys in JSON documents are often unique in an object whereas different sub-elements of an XML element can be of same tags. Additionally, JSON has important performance and resource utilization advantages over XML\cite{comp}, which makes the JSON-like document databases particular interesting.

In recent years some document databases like MongoDB\cite{mongodb}, CouchDB\cite{couchdb}, OrientDB\cite{orientdb} and RavenDB\cite{ravendb} have been developed and are becoming popular in practice. With the wide spread use of document databases, how to query JSON-like documents has become an important issue. Existing document databases usually provide lightweight query interfaces to satisfy basic query needs. For instance, MongoDB uses certain APIs, such as the ``\emph{find()}'' function with the keys and constraints as the arguments, to extract the corresponding elements from source documents in the runtime. CouchDB develops a view-based query interface where the predefined views are key-value pairs generated by underlying map/reduce functions. OrientDB supports a simple subset of SQL to extract the document fragments in collection of records, besides its native java-based query API. RavenDB presents its queries with LINQ programs on .NET framework, using the classes developed on indices. Although the above databases allows simple data extraction from documents, their query APIs are not expressive enough for complex requests. For example, to extract multiple interdependent data elements or to construct a specified data structure, users often have to resort to host programming languages. Therefore, a full-fledged query language is necessary to facilitate presenting complex document queries.

Some researchers have begun to study query languages for JSON-like documents recently. Jaql\cite{jaql} is a functional query language that enables users with function pipelines to extract, filter, join and group JSON data. It enables computation on data collections with the \emph{transform} command and allows some simple structure manipulation with the commands like \emph{expand}.  JSONiq\cite{jsoniq} is a query language which adopts a subset of XQuery syntax and semantics to query JSON documents, by treating JSON documents as simple XML documents. Similarly, some studies \cite{jpath,jsonpath,jsonselect} also employ XPath-like expressions to extract JSON documents in a navigational way, but they have no   coherent and complete filtering and constructing mechanism and thus cannot be called a full-fledged language. UnQL\cite{unql} is a SQL-like JSON query language which enables selecting and manipulating object elements in data hierarchy with navigation operators. These languages are more expressive than the aforementioned query interfaces, but they are not expressive and declarative enough to present complex queries.  More specifically, since they often use navigation operators on documents to extract homogeneous data as plain tuples, it is not easy for them to extract multiple interdependent elements with inherent structure, to handle heterogeneous data and to construct hierarchical document structure. These common requests have to be carried out by multiple extraction commands, nested loop statements and other specific mechanisms. It requires the users to consider the detailed procedures of queries themselves and to be skillful in programming.

In this paper, we present our preliminary study on a  pattern-based query language named JPQ (standing for JSON-like Pattern Query).  JPQ is a functional language which is inspired by our former study on declarative XML queries\cite{xtq}. It adopts expressive patterns with coherent mechanisms to extract multiple document data elements simultaneously and construct document fragments in a flexible structure, and thus enabling complex document queries be presented in a declarative way. Our work makes the following contributions: a) Although some pattern-based query languages were once studied on XML, to our knowledge our work is the first attempt to use tree-like patterns in querying JSON style document databases. b) In comparison with existing studies on document query, JPQ focuses on the expressiveness of the query language and adopts various kinds of data extraction patterns to facilitate presenting complex document queries which are useful in practice.  c) JPQ deploys a deductive rewriting mechanism to specify data transformation requests, which allows complex data construction to be presented declaratively.

The remainder of the paper is organized as follows. In order to present our query language, we first introduce a simple model of JSON-like documents in Section 2, together with a brief discussion on the schema-less feature and its effects on document database queries. Section 3 introduces the JPQ language with certain examples demonstrating its usage and features. Section 4 compares JPQ and other  document query languages or interfaces and concludes the paper.

\section{A Simplified Model for Document Query}

JPQ adopts a simplified model of JSON-like documents named JHM, standing for JSON-like Hierarchy Model. JHM uses the following grammar to specify the data structure of key-value document fragments in JSON style. As the grammar shows, a JHM document fragment is represented as a value which can be an atomic one like a string or a number, an array of values, or an object comprising one or more key-value pairs. A JHM document often exhibits a data hierarchy as the values in objects or arrays are expanded. JHM requires the string-based keys to be unique in an object, as document database usually does.

\begin{small}\begin{verbatim}
<value> ::= <atom> | <array> | <object>
<atom> ::= <string> | <number> | <boolean> | empty | ...
<array> ::= [ <value> (; <value>)* ]
<object> ::= {<keyvalue> (, <keyvalue>)* }
<keyvalue> ::= <string>:<value>
\end{verbatim}\end{small}

Although JSON format is evolved from the object-oriented paradigm, document data as JHM specifies are different from conventional object data. The fundamental distinction is that JHM adopts no compound data type and, further, no predefined data schema. This schema-less property essentially affects document queries in several ways. Firstly, it disables randomly accessing an arbitrary part of an object, thus a certain enumeration mechanism is required to parse the object and get the details. Secondly, it does not use labels with a predefined semantics, thus querying key strings becomes an important way to find expected information, because the key strings with the self-described semantics play the role of semantic labels of the associating values. Finally, it supports the heterogeneity of array elements which are often assumed to convey similar information, thus users often need to manually align heterogeneous array elements to produce homogeneous results.

The following is a document fragment \emph{univ} illustrating a scenario of a university human resource information system where the information of the presidents, the schools and the faculties is stored in a hierarchical document.

\begin{small}\begin{verbatim}
{ "president":{"ID":"0001", "last name":"Li", "first name":"XH",
               "email":"xxli@123.edu"},
  "executive-vice-president":{"ID":"0002","last name":"Feng",
               "firstname":"YM", "email":xxfeng@123.edu"},
  "vice-presidents":[{"ID":"0003","surname":"Zhou","givenname":"CB"};...]
  "schools":[{"name":"Computer School", "dean":{"ID":"0011", ...},
              "faculty":[{"ID":"0001", "first name":"Li",...,
                          "email":"xxli@cs.123.edu:"};
                        ...]},
             ... ],
  ... }                 (univ)
\end{verbatim}\end{small}

%The document fragment \emph{univ} illustrates a scenario of a university human resource information system For simplicity, we assume that each person object contains a key ``ID'' with a unique identity value.

We consider the following three queries on the human resource information: a) find all presidents' personal information; b) for each president find the school he/she belongs to; and c) for each faculty member list the schools in which he/she has an occupation. These queries are common in practical information system. However, if a user tries to present the queries using existing document query interfaces or languages, he might have to spend a lot of time to write tedious programs to extract interdependent data elements, to handle heterogeneity and to transform hierarchical data structure. So here comes JPQ.

%The JPQ language is developed to facilitate the tasks.

%As the fragment shows, the document comprises the 2-level human resource information of a university. values of presidents including a ``president'' object, an ``executive-vice-president'' object and a ``vice-presidents'' array locate in the first level; in the second level, for each subordinate school of the university, its dean and faculties are represented as an object and an array respectively. and the name information whose keys, such as ``first name'' or ``firstname'',  often vary for different people. Some additional information such as ``email'' or ``phone-number'' are optional for people value.
%
%

%
% For example, a sample fragment about certain book information is listed below.
%
%\begin{small}\begin{verbatim}
%{ book : [ {
%    year : "1999", title : "Term Rewriting and All That",
%    author : [
%     {last:"Franz", first:"Baader", email:"baader@tcs.inf.tu-dresden.de"},
%     {surname:"Tobias", givenname:"Nipkow"} ],
%    editor : {last:"David", first:"Tranah", email:"dtranah@cup.cam.ac.uk"},
%    price:"19.99" },
%   ...] }                              (sampledoc)
%\end{verbatim}\end{small}
%
%

\section{The JPQ Language}
\subsection{Synopsis of JPQ}

JPQ is a declarative query language which deploys hierarchical patterns to specify queries on JHM documents. It stems from our previous work on XML query but is considerately modified to cater for the practical document query requests mentioned above. A common JPQ program is composed of  the \textbf{from}, the \textbf{construct} and the \textbf{where} clauses to present data extraction, construction and filtering respectively. The synopsis of the JPQ grammar is listed below.
\begin{verbatim}
<query> ::= from <doc> <extraction-pattern>
                 (,<doc> <extraction-pattern>)*
            construct <construction-pattern>
            where <conditions>
\end{verbatim}

A JPQ query works as follows. The \emph{from} clause in a JPQ program uses one or more extraction statements, i.e., an extraction pattern following a source document, to specify extraction requests. The data elements in the document are extracted if their surrounding elements and contexts match the corresponding parts of the pattern. These data elements are naturally organized following the structure derived from the extraction pattern. This structure is coherently restructured and used in the construction pattern of the \emph{construct} clause to transform the extracted data elements and generate the output results. Additionally, if there is a \emph{where} clause,  the extracted data elements should be filtered by the conditions specified in it before the construction of the final results.

%To satisfy the requests of data transformation, matching terms are often restructured and then used to form a construction pattern, indicating the structure of the transformed data.

%Instead of listing the syntactical and semantical specification of the language, in the following subsections, we use some typical query examples on the university document ``univ'' to show the essential features and usages of the JPQ language.

\subsection{Data Extraction}

Extraction patterns in JPQ include key-value patterns and value patterns which are to respectively match the key-value pairs and the values in JHM documents. We use  \emph{p$_{k}$} and \emph{p$_{v}$} to denote the key-value pattern and the value pattern,  use \emph{v} to denote the variables, use \emph{r$_{s}$} and \emph{r$_{v}$} to denote the string predicate and the value predicate, and list the abstract syntax of the extraction patterns as below:\vspace{5pt}\\
\emph{
\indent p$_{k}$ ::= v:p$_{v}$ $|$ r$_{s}$:p$_{v}$ $|$ (v r$_{s}$):p$_{v}$ $|$ *:p$_{v}$ $|$ p$_{k}$$|$p$_{k}$\\
%\indent p$_{k}$ ::= k:p$_{v}$ $|$ p$_{k}$$|$p$_{k}$\\
%\indent k ::=  v $|$ r$_{s}$ $|$ v r$_{s}$ $|$ *\\
\indent p$_{v}$ ::= v $|$ r$_{v}$ $|$ * $|$ \{p$_{k}$, ..., p$_{k}$\} $|$ [p$_{v}$] $|$ $<$p$_{v}$,p$_{v}$$>$ $|$ p$_{v}|$p$_{v}$ $|$ /p$_{k}$ $|$ //p$_{v}$
}
\vspace{5pt}

As the syntax rules show, an extraction pattern is often structurally composed of the variables, the predicates and the wildcard ``*'' with the structural operators like ``:'', ``\{\}'' or ``[ ]'' and the logical operators like ``$<,>$'' or ``$|$''. The variables and the predicates for the key string or value are to test whether a document fragment, i.e., a key or a value, satisfies the restrictions indicated by the pattern, and if it does, to bind the fragment to the variable (if there is one) in the pattern. For example, to match a key-value pair ``\textsl{{`executive-vice-president'}:v}'' with a key-value pattern ``\textsl{\$k `?president?':*}'', the key \textsl{``executive-vice-president''} is tested by the string predicate ``\textsl{?president?}'' for the existence of a substring ``\textsl{president}'', and \textsl{\$k} would be bound to the key, denoted as the matching pair ``\textsl{\$k$\mapsto${`executive-vice-president'}}''. The value \textsl{v} is ignored as it is matched with the wildcard  ``*''. On the other hand, matching the element with the key-value pattern ``\textsl{`?president?':\$x}'' would result in a matching pair ``\textsl{\$x$\mapsto$v}''.

For an extraction pattern which can bind various parts of a document fragment to one or more variables, the matching pairs like ``\textsl{\$x$\mapsto$v}'' are organized structurally according to the composite structure of the pattern. JPQ adopts a logical way to compose the matching pairs by introducing three kinds of structures namely \textbf{tuple}, \textbf{array} and \textbf{option}.

A tuple of the form \textsl{(r$_{1}$,\dots,r$_{n}$)} conjunctively combines the subordinate matching results \textsl{r$_{1}$,\dots, r$_{n}$}. It is generated in matching a fragment with a multi-variable pattern, such as a definitive key-value pattern, an object pattern or a conjunctive pattern, where the variables are considered to be conjunctively associated. JPQ also introduces the conjunctive pattern \textsl{$<$p$_{1}$,...,p$_{n}$$>$} for the queries where a value is to be respectively matched with the patterns \textsl{p$_{1}$,...,p$_{n}$} and the results are combined as a tuple. Here the pattern \emph{p$_{i}$} can also be a predicate required to be satisfied by the value.
For example, matching the aforementioned key-value with the pattern ``\textsl{\$k`?president?':$<$\$p,\{`last name': $<$\$l,`F?'$>$\}$>$}'' would generate a tuple ``\textsl{(\$k$\mapsto$`executive-vice-president',\$p$\mapsto$v,\$l$\mapsto$`Feng')}''.

%matching the aforementioned key-value pair with the pattern ``\textsl{\$k`?president':\$x}'' would generate a tuple of matching pairs ``\textsl{(\$k$\mapsto$`executive-vice-president', \$x$\mapsto$v)}'';
%For example, the pattern ``\textsl{}'' is used to match a value of president whose last name begins with ``L'' and generate a tuple such as ``\textsl{(\$p$\mapsto$v, \$l$\mapsto$`Li')}''.

An array of the form \textsl{[r$_{1}$;\dots;r$_{n}$]} is an ordered list of the results \textsl{r$_{1}$,\dots,r$_{n}$}, which derives from matching fragments with an array pattern or an enumeration pattern. For example, matching the ``\textsl{`vice-presidents':[v$_{1}$;\dots;v$_{n}$]}'' in \textsl{univ} (where \textsl{v$_{1}$,...,v$_{n}$} denotes the values in the array) with the pattern ``\textsl{`?president?':[\$p]}'' will result in the array ``\textsl{[\$p$\mapsto$v$_{1}$;\dots;\$p$\mapsto$v$_{n}$]}''.

JPQ introduces the enumeration patterns for generating an array of matching results from an arbitrary fragment, borrowing the convenient path operators ``\textsl{/}'' and ``\textsl{//}'' from XML query languages. These patterns are used to parse and enumerate the content of a fragment when its inner structure is unknown, as previously mentioned. The children  pattern \textsl{/p} is to match an object value by iterating its elements and matching them with the pattern \textsl{p} . The matching results of \textsl{p} would form an array following the document order. For example, matching \textsl{univ} with the pattern ``\textsl{/\$r `?president?':*}'' will result in the array ``\textsl{[\$r$\mapsto$`president'; \$r$\mapsto$`executive-vice-president'; \$r$\mapsto$`vice-presidents']}''.
The descendants pattern \textsl{//p} is to recursively find all the values matching the pattern \textsl{p} under current context and the sub-contexts.
The matching results would form an array following the preorder-traversal order.

An option derives from matching a fragment with an optional pattern \textsl{p$_{1}$$|$p$_{2}$\dots $|$p$_{n}$} which provides multiple choices of pattern matching. The fragment would be disjunctively matched with the patterns \textsl{p$_{1}$, \dots, p$_{n}$}, and the results would be filtered by the conditions in the \textsl{where} clause if such conditions exists. After that, the first valid result of p$_{i}$ would be used as the result of the whole pattern.
Option patterns are useful in handling heterogeneity. For example, matching a person value of univ with the pattern ``\textsl{`first name':\$f$_{1}$$|$`firstname':\$f$_{2}$$|$`given name':\$g}'' will return his/her first name or given name, and thus heterogeneous person names can be unified and processed homogeneously; on the other hand, matching a name value with the pattern ``\textsl{$<$\$n1, `L?'$>|<$\$n2,'W?'$>$}'' will bind the value to different variables depending its initial letter, and thus homogeneous data can be separated and handled heterogeneously.

Using the extraction patterns described above, JPQ can flexibly and expressively extract and organize the elements.
%which nicely satisfies the requirements for presenting document data queries as listed in the previous section.
For example, to find the schools where a president works as a faculty member, we can use the following extraction pattern to find the related information.
\begin{small}\begin{verbatim}
doc("univ") </$r"?president?":(<$p1,{"ID":$id1}>|[<$p2,{"ID":$id2}>]),
             {"schools":[{"name":$n, "faculty":[{"ID":$id3}]}]}>
\end{verbatim}\end{small}
This extraction pattern is a conjunctive one composed of the enumeration pattern ``\textsl{/\$r`?president?':...}'' and the object pattern ``\textsl{\{`schools':...\}}''. In matching the document with the former pattern, all the key-value pairs in the object would be enumerated and matched with the key-value pattern. The key-value pairs whose key contains the substring ``\textsl{president}'' would be bound to \textsl{\$r} and its value, denoted as \textsl{v$_{1}$}, \textsl{v$_{2}$}, \dots,  would be matched with the option pattern ``\textsl{$<$\$p1...$>|$[$<$\$p2...$>$]}''. In matching the document with the latter pattern, the array value of the ``\textsl{schools}'' would be matched with the succeeding array pattern to extract each school's name and IDs of its faculty members. Part of the matching result is listed as follows:
\begin{small}\begin{verbatim}
( [ ($r -> "president", $p1 -> v1, $id1 -> "0001");
    ($r -> "executive-vice-president", $p1 -> v2, $id1 -> "0002");
    ($r -> "vice-presidents", [($p2 -> v3, $id2 -> "0003"); ... ])   ],
  [ ($n -> "Computer School", [$id3 -> "0001"; ...]); ...  ] )
\end{verbatim}\end{small}

\subsection{Data Construction}
%
%
%JPQ uses an expression named \textbf{matching term} to specify the structure of the matching result. A matching term is composed of the variables in the result and the operators like ``(,)'', ``[ ]'' and ``$|$'', and thus is also called a tuple, array or option. For example, a tuple result ``\textsl{(\$x$\mapsto$v$_{1}$,\$y$\mapsto$v$_{2}$)}'' is of the tuple term ``\textsl{(\$x,\$y)}'', and an array ``\textsl{[\$x$\mapsto$v$_{1}$;\$y$\mapsto$v$_{2}$]}'' is of the array term ``\textsl{[\$x$|$\$y]}''. Matching terms can exactly represent the structure of the extracted data and can be directly derived from the extraction patterns. For the above example pattern, its matching term is ``\textsl{([(\$r,(\$p1,\$id1)$|$[(\$p2,\$id2)])], [(\$n, [\$id3])])}''. The matching term of an extraction pattern \textsl{p} is often denoted as a function ``\textsl{mt(p)}''. The formal definition of matching terms would be specified in the next subsection.

Data construction in JPQ is presented with construction patterns which are essentially the function invocations on the \textbf{matching terms}. A matching term is an expression to specify the (transformed) structure of the matching results of an extraction pattern. The abstract syntax of matching term is\vspace{5pt} \\
\indent \indent \indent \emph{t ::= v $|$ (t,t) $|$ t$|$t $|$ [t]$_{t}$ $|$ \^{}[t]$_{t}$ $|$ t\%}\vspace{5pt} \\ % $|$ $\epsilon$}\\
As the syntax shows, a matching term is often composed of the variables in the result and the operators like ``(,)'', ``[ ]'', ``$|$'' and ``\%'', and thus is called a tuple, array, option or distinct term. For the array term \textsl{[t]$_{t'}$}, \textsl{t'} is named the \textbf{index term} of \textsl{t} and \textsl{t} can usually be specified as a \textsl{(t',t'')}. It is required that in the array of matching results with the term \textsl{[t]$_{t'}$}, each item of \textsl{t} can be identified by an item of \textsl{t'}. Generally, the element of an array of the original matching results should be indexed by itself, that is, its matching term should be like \textsl{[t]$_{t}$}.
 For the distinct term \textsl{t\%}, it denotes a distinct value in a given array, which is introduced in detail later.
For example, a tuple result ``\textsl{(\$x$\mapsto$v$_{1}$,\$y$\mapsto$v$_{2}$)}'' is of the tuple term ``\textsl{(\$x,\$y)}'', and an array ``\textsl{[\$x$\mapsto$v$_{1}$;\$y$\mapsto$v$_{2}$;\$x$\mapsto$v$_{3}$]}'' is of the array term ``\textsl{[\$x$|$\$y]$_{\$x|\$y}$}''. The matching term of an extraction pattern \textsl{p} is often denoted as a function ``\textsl{mt(p)}''.

%The argument matching term of a construction pattern is named its \textsl{backbone}.
A construction pattern can be specified as a normal function invocation like \textsl{fun(t)} or, more often, by embedding certain constant values (such as key strings or values) and necessary notations (such as ``:'' and ``\{\}'') into the argument matching term. In the runtime the construction pattern can generate a valid JHM fragment by instantiating the variables with the values they are bound to. For example, the construction pattern ``\textsl{`people':[\{`surname':\$n\}]}'' has the argument term ``\textsl{[\$n]}'', and it would generate the fragment ``\textsl{`people':[\{`surname':`Li'\};\{`surname': `Gu'\}]}'' given the matching result ``\textsl{[\$n$\mapsto$`Li'; \$n$\mapsto$`Gu']}''.
Besides, function invocations on the subordinate matching terms are allowed to occur in a construction pattern, which makes data construction expressive and compact.

%\begin{example} List the length and the surnames or last names of the vice-presidents array.
%\begin{small}\begin{verbatim}
%from doc("univ"){"vice-presidents":[$p]}
%construct {"count":count([$p]), "surnames":[$p.("surname"|"last name")]}
%\end{verbatim}\end{small}
%\end{example}
%This example would generate an object value composed of two key-value pairs by two function invocations: ``\textsl{count([\$p])}'' would return the array length, and ``\textsl{\$p.(`surname'$|$`last name')}'' would return the values associating with the key ``\textsl{surname}'' or ``\textsl{last name}'' in the object value bound to \textsl{\$p}.

A more common request on data construction is to transform the original extracted data to be of a specified structure rather than follow the original structure in the source document. JPQ can declaratively present the data transformation by employing a deductive restructuring mechanism of matching terms. This mechanism deploys a set of restructuring rules, which constitute a term rewriting system, to indicate the restructuring of matching terms and accordingly direct the transformation of matching results. This method has been successfully utilized in our studies on XML query\cite{xtq}. In JPQ we substantially modify it to be simple and powerful enough for document database query.
The restructuring rules of the matching terms are listed in Fig.1.

\begin{figure}[h]
\begin{center}
\begin{tabular}{|p{\textwidth}|}
  \hline
%\textbf{Auxiliary rules} \\
\hspace{10pt}(t$_{1}$,\dots,  t$_{i}$, t$_{i+1}$,\dots, t$_{n}$)$\hookrightarrow$ (t$_{1}$,\dots, t$_{i+1}$, t$_{i}$,\dots, t$_{n}$)   (tuple-commutation)\\
\hspace{10pt}(t$_{1}$,\dots,  t$_{j}$, t$_{j+1}$,\dots, t$_{n}$)$\hookrightarrow$ (t$_{1}$,\dots, t$_{j}$, (t$_{j+1}$,\dots, t$_{n}$))    (tuple-association) \\
\hspace{10pt}(t$_{1}$$|$\dots$|$t$_{i}$$|$t$_{i+1}$$|$\dots$|$\dots$|$t$_{n}$)$\hookrightarrow$ (t$_{1}$$|$\dots$|$t$_{i+1}$$|$t$_{i}$$|$\dots$|$ t$_{n}$)       (option-commutation)\\
\hspace{10pt}(t$_{1}$$|$\dots$|$ t$_{j}$$|$t$_{j+1}$$|$\dots$|$t$_{n}$)$\hookrightarrow$ (t$_{1}$$|$\dots$|$t$_{j}$$|$(t$_{j+1}$$|$\dots$|$t$_{n}$))     (option-association)\\
%\hline
%\textbf{Reducing and Duplicating rules}\\
\hspace{10pt} t $\hookrightarrow$ (t, t) (tuple-duplication)
%\hspace{10pt}(t, t')$\hookrightarrow$ t / t'  if var(t)$\cap$ var(t') = $\phi$ (tuple-hiding)\\
%\hspace{10pt}(t $|$ t')$\hookrightarrow$ t  (option-hiding)\\
%\hline
%\textbf{Distributing rules} \\
%%(t, \{t\}) $\hookrightarrow$ $\varepsilon$ (tpl-grp-invalid) \\
%(t, \{t'\}) $\hookrightarrow$ \{(t, t')\}$_{\{t'\}}$   if  var(t)$\cap$ var(t') = $\phi$ (grp-distr1) \\
%(t, (t' $||$ t'')) $\hookrightarrow$ ((t, t') $||$ (t, t''))\\
% \hspace{20pt} if var(t)$\cap$ (var(t')$\cup$ var(t'')) = $\phi$
%(enum-distr)\\
%\textbf{Grouping rules}\\
%\hspace{10pt}[t]$_{i}$ $|$ [t']$_{i'}$ $\hookrightarrow$ [t$|$t']$_{i|i'}$ (array-option-expand)
\hspace{10pt}[t]$_{i}$  $\hookrightarrow$ \^{}[t]$_{i}$ (array-flattening) \\
\hspace{10pt}(t, t' $|$ t'' ) $\hookrightarrow$  (t, t') $|$ (t, t'')  (option-tuple-distribution)\\
\hspace{10pt}(t, [t']$_{i}$) $\hookrightarrow$ [(t, t')]$_{i}$   if  var(t)$\cap$ var(t') = $\phi$, [t']$_{i}$ is not folded array. \\
\hspace{40pt}(array-tuple-distribution) \\
%\hspace{20pt}
%\{\^{}\{t\}$_{ql}$\}$_{ql'}$  $\hookrightarrow$ \{t\}$_{ql': ql}$  (grpflatten--red)\\
%\{t $||$ t'\}$_{ql}$ $\hookrightarrow$ $<$\{t\}$_{ql}$ $||$ \{t'\}$_{ql}$$>$ (grp-split) \hspace{15pt}
\hspace{10pt}[(t, t')]$_{i}$ $\hookrightarrow$ [([(t, t')]$_{i}$, t\%)]$_{[t\%]}$ (array-tpl-folding)   \\
%\hline
%\textbf{Option rules} \\
%\hspace{10pt}t $|$ t  $\hookrightarrow$  t  (option-idemoptence) \\
%\hspace{10pt}([t]$_{[q]}$ $|$[t']$_{[q']}$)  $\hookrightarrow$ [t $|$ t']$_{[q|q']}$ (option-array-alter) \\
\hline
\end{tabular}
\caption{Restructuring rules of matching term}
\end{center}
\end{figure}

These restructuring rules specify the atomic steps of data transformation indicated by restructuring matching terms. For two matching terms \textsl{t} and \textsl{t'}, we say \textsl{t} is valid (with respect to \textsl{t'}), if \textsl{t'} can be restructured to \textsl{t} by applying the restructuring rules sequentially. For the construction pattern in the construct clause of a JPQ program, it is required that the backbone of the construction pattern is valid with respect to the matching term derived from the extraction pattern. For the valid data transformation indicated by a valid restructured matching term, its restructuring route can be inferred by the deductive mechanism of the rewriting system. Therefore, the requests of data transformation in JPQ can be presented declaratively.
Due to space restrictions, we would demonstrate the usage and semantics of the rules with some typical examples. For the formal details and theoretical properties of term restructuring and data transformation, please refer to our technique report\cite{report}.

The commutation and the association rules of the tuples and the options are used to rearrange the order of the subordinate terms in the construction pattern. %and accordingly affect the order of the subordinate elements.
The tuple-duplication rule is used to duplicate data elements to satisfy different construction requests. %As shown in Example 1, the construction pattern is built on the matching term ``\textsl{([\$p],[\$p])}'' which is the duplication of the original term ``\textsl{[\$p]}''. %The former array is used to count the number of members, and the latter one is used to get the members' surnames or last names.
%A hiding construction pattern is introduced in JPQ to hide a part of a tuple in the form of ``\textsl{t hid t'}'' whose argument term is \textsl{(t,t')}. \textsl{t hid t'} indicates that only the exposed part \textsl{t} be used in data construction.. Especially, a hiding  pattern \textsl{t hid t'} can be simplified as \textsl{t} if it incurs no misunderstanding.
%tuple \textsl{p/q} in a construction pattern can be simplified as \textsl{p} by omitting the term \textsl{q} if it incurs no misunderstanding, otherwise it should be explicitly signified as``\textsl{p hid q}''.
%Hiding part of the values in the extracted data or duplicated data is often necessary in data construction. JPQ allows a tuple or an option to hide part of its subordinate patterns, as shown in the two hiding rules,

%\begin{example} Separate the surnames and the last names of the vice-presidents array into two arrays.
%\begin{small}\begin{verbatim}
%from doc("univ"){"vice-presidents":[{"last name":$l|"surname":$s}]}
%construct {"last names":[$l], "surnames":[$s]}
%\end{verbatim}\end{small}
%\end{example}
%In this example, the original matching term ``\textsl{[\$l$|$\$s]}'' is firstly duplicated as ``\textsl{([\$l$|$\$s],[\$l$|$\$s])}'' and then reduced to ``\textsl{([\$l],[\$s])}'' by hiding the sub-patterns in the options accordingly.

As the data elements in a JHM document are often hierarchically organized, it is natural for the extracted data to contain nested arrays. It is a common request to flatten a nested array, that is, to make the elements of the inner array to be the direct elements of the outer array. JPQ introduces a flattened array term \textsl{\^{}[t]$_{i}$} and specifies the array-flattening rule to transform an array \textsl{[t]$_{i}$} to a flattened array ``\textsl{\^{}[t]$_{i}$}'' if \textsl{[t]$_{i}$} is a member of an outer array like \textsl{[[t]$_{i}$]$_{i'}$}. After the transformation, the elements in the inner arrays of \textsl{[[t]$_{i}$]$_{i'}$} would be released and directly belong to the outer array.

%\begin{example}Generate an array containing all the faculty member values of the university.
%\begin{small}\begin{verbatim}
%from doc("univ"){"schools":[{"faculty":[$f]}]}
%construct {"faculty-members":[^[$f]]}
%\end{verbatim}\end{small}
%\end{example}
%In this example, for a nested array like ``\textsl{[[\$f$\mapsto$`Li';\$f$\mapsto$`Wu'];[\$f$\mapsto$`Liu'];
%[\$f$\mapsto$`Zhu';\$f$\mapsto$`Gu']]}'', the flattened array term \textsl{[\^{}[\$f]]} would generate the result ``\textsl{[\$f$\mapsto$`Li';\$f$\mapsto$`Wu'; \$f$\mapsto$`Liu';\$f$\mapsto$`Zhu';\$f$\mapsto$`Gu']}''.

%distribution
JPQ allows a tuple containing an option or array to be restructured to an option or an array, as shown in the two distribution rules. This kind of transformation is named distribution because it works like a distribution law in algebra. An option distribution is to restructure the tuple \textsl{(t, t$_{1}$$|$t$_{2}$)} to the option \textsl{(t,t$_{1}$)$|$(t,t$_{2}$)} so that the values bound to \textsl{t} can be embedded in different construction pattern depending on its associating pattern \textsl{t$_{1}$} or \textsl{t$_{2}$}.
An array distribution is to restructure the tuple \textsl{(t,[t']$_{i}$)} to the array \textsl{[(t,t')]$_{i}$}, that is, to distribute the values of \textsl{t'} to be coupled with the value of \textsl{t} and thus form a new array \textsl{[(t,t')]}. In the new array \textsl{[(t,t')]}, each element corresponds to an element in the array \textsl{[t']$_{i}$} and thus the restructured array can be indexed by the original index term \textsl{i}. The information of the index array of an array is very helpful for inferring the provenance of array distribution. In a construction pattern an array \textsl{[t]$_{i}$} is presented as \textsl{[t] \textbf{groupby} i} whereas it is often simplified as \textsl{[t]} if the index term \textsl{i} can be inferred.

\begin{example}Find all the presidents and their associating object values, and generate a ``presidents'' array containing each president's role and the corresponding object value.
\begin{small}\begin{verbatim}
from doc("univ")/($r"?president?"):(<$po,{}>)|[$pa])
construct {"presidents":[{"role":$r,"info":$po}|^[{"role":$r,"info":$pa}]]}
\end{verbatim}\end{small}
\end{example}
As shown in Example 1, the key-value pairs of the presidents are enumerated,  the object value would be bound to ``\textsl{\$po}'' directly and the array value would be further matched with ``\textsl{[\$pa]}''. In the construction pattern,  the raw term ``\textsl{(\$r,\$po$|$[\$pa])}'' is restructured to the term ``\textsl{(\$r,\$po)$|$(\$r,[\$pa])}'' in which \textsl{\$r} would be used in different ways. Further, the tuple ``\textsl{(\$r,[\$pa])}'' is restructured to the array ``\textsl{[(\$r,\$pa)]}'' and then flattened as ``\textsl{\^{}[(\$r,\$pa)]}''. Therefore, in the construction result the  presidents' objects and the vice-presidents' array members would be merged into a homogeneous array.

Besides indicating the restructuring provenance, an index array can also be used to reorder the elements in the array it refers to. JPQ allows the matching term in an index array to be used as function argument and introduces the ``\textbf{asc}'' and ``\textbf{desc}'' suffix to sort the function values of the elements in the index array. The elements in the value array are accordingly reordered.

%value grouping
Grouping elements by a set of specified index is a common request in data query. JPQ implements this request with folded arrays and the array-tuple-folding rule. The tuple elements in the array \textsl{[(t,t')]$_{i}$} can be grouped as an array of equivalence classes on the values of the term \textsl{t} which is denoted as the folded array \textsl{[([(t,t')]$_{i}$,t\%)]}. Naturally, the folded array has the index term \textsl{t\%}, indicating the distinct values of the array \textsl{[t]}. %Especially, the number of the elements for a value of \textsl{t\%} in the array \textsl{[t]} can be get by the function ``\textsl{numberof(t\%)}''.
In the construction patterns, the index pattern \textsl{t\%} is also presented as ``\textsl{groupby t\%}'' and is often omitted if \textsl{t\%} is used as an indication.

\begin{example}For each faculty member list the schools in which he/she has an occupation, order the faculty members in the ascending order of their IDs.
\begin{small}\begin{verbatim}
from doc("univ"){"schools":[{"name":$n, "faculty":[{"ID":$id}]}]}
construct {"faculty":[{"ID":^[$id]%, ["school":$n]}] groupby ^[id]% asc}
\end{verbatim}\end{small}
\end{example}
As shown in Example 2, the nested array denoted by ``\textsl{[(\$n,[\$id])]}'' is firstly flattened as ``\textsl{[(\$n,\^{}[\$id])]}'' and then be classified as ``\textsl{[([(\$n,\^{}[\$id])],\^{}[\$id]\%)]}'' which is used in the construction pattern by hiding the term \textsl{\^{}[\$id]} in the inner arrays. The index term \textsl{\^{}[id]\%} has the ordering suffix ``\textsl{asc}'', indicating to order the array elements by the ascending order of ID.
%Additionally, if the user wants to order the array elements by the number of the schools a faculty member works for, the groupby suffix in the above construction pattern can be changed as ``\textsl{groupby numberof(\^{}[id]\%) asc}''.

\subsection{Data Filtering}

JPQ uses predicate conditions and compound conditions in the \emph{where} clause to filter out the unwanted data elements. Predicate conditions are essentially predicate function invocations to filter the values bound to the variables in the argument term. A predicate condition can be a simple predicate function invocation, a quantified condition or a composite condition combining subordinate conditions with the boolean connectives ``\textbf{and}'', ``\textbf{or}'' and ``\textbf{not}''.  The simple predicate function invocation is in the canonical form as \textsl{fun(p)}, or is in the infix form like ``\textsl{\$a1 = \$a2}''  for some binary functions.

\begin{example}Find the pairs of names of the schools which share common faculty member.
\begin{small}\begin{verbatim}
from doc("univ"){"schools":<[{"name":$n1, "faculty":[{"ID":$id1}]}],
                            [{"name":$n2, "faculty":[{"ID":$id2}]}]>}
construct {"result":[(^[{"school1":$n1, "school2":$n2}])]}
where not($n1 = $n2) and $id1=$id2
\end{verbatim}\end{small}
\end{example}
In this example, a self-join on the school arrays is established with the composite condition. As the two simple conditions respectively have the arguments ``\textsl{(\$n1,\$n2)}'' and ``\textsl{(\$id1,\$id2)}'',  the composite condition's argument is the tuple term ``\textsl{(\$n1,\$n2,\$id1,\$id2)}''.

Quantified conditions are the ones with the quantifiers ``\textsl{foreach}'' or ``\textsl{forsome}'', in the form of ``\textsl{quantifier mterm in array; predicate-condition}'', indicating that each (or some) of the element(s) bound to the term in the array should satisfy the predicate condition in the quantified condition body.  The quantified term ``\textsl{p in [p]}'' can be simplified as \textsl{p}.
\begin{example} Find the schools whose faculty amount is larger than 100 or each member has an email address.
\begin{small}\begin{verbatim}
from doc("univ"){"schools":[($s {"faculty":[$f]})]}
construct "result":[$s]
where count[$f]>100 or (foreach $f; notnull($f."email"))
\end{verbatim}\end{small}
\end{example}
A quantified condition's argument is the array it ranges in rather than the array elements. That is, the two subordinate conditions all have the argument term ``\textsl{[\$f]}'' and thus the composite condition also has the argument ``\textsl{[\$f]}''.

Since different predicate conditions are used to filter the data in different part of the data structure, JPQ adopts compound conditions, the predicate conditions combined with the connectives ``\textbf{par}'' and ``\textbf{with}'', to form a consistent global requirement of filtering the extracted data. A compound condition \textsl{c$_{1}$ par c$_{2}$} is a disjunctive condition where \textsl{c$_{1}$} and \textsl{c$_{2}$} are two conditions filtering values of different parts of an option. In this condition the two conditions work in parallel and the filtering results are merged as the final result of the whole condition.

%par condition
\begin{example}For each president find the schools he/she is also a faculty member.
\begin{small}\begin{verbatim}
from doc("univ")</"?president?":(($p1{"ID":$id1})|[($p2{"ID":$id2})]),
                 {"schools":[{"name":$n, "faculty":[{"ID":$id3}]}]}>
construct {"results":[^[{"president":$p1,"school":$n}]|
                      ^[^[{"president":$p2,"school":$n}]]]}
where id1=id3 par id2=id3
\end{verbatim}\end{small}
\end{example}
In this example, each of the two predicate conditions specifies one branch of the option term ``\textsl{(\$id1,\$id3)$|$(\$id2,\$id3)}''. Therefore, we need to combine them with the ``par'' operator to gather the results filtered by two conditions. A common illusion is to use the operator ``or'' instead of ``par''. However, using ``or'' means that the composite condition can be treated as a predicate with the argument term ``\textsl{(\$id1,\$id2,\$id3)}'', which is not a valid function in the JPQ language.

A compound condition \textsl{c$_{1}$ with c$_{2}$} is used to filter hierarchical data elements. In the runtime, the predicate condition \textsl{c$_{2}$} works on the result data elements filtered by the predicate \textsl{c$_{1}$}. This condition differs from ``\textsl{c$_{1}$ and c$_{2}$}'' in that it is processed sequentially rather than commutatively, which indicates a bottom-up order on data hierarchy.
%with condition
\begin{example}Find the schools whose faculty contains at least 100 members who have an email address with an ``edu'' suffix.
\begin{small}\begin{verbatim}
from doc("univ"){"schools":[{"name":$n, "faculty":[{"email":$m}]}]}
construct {"result":[{"school":$n}]}
where endWith($m,"edu") with count([{$m}])>=100
\end{verbatim}\end{small}
\end{example}
In this example, the condition to filter the faculty members and the one to filter the arrays of the faculties work at different levels in the data hierarchy, and thus a bottom-up order to process the data filtering is required.

Data filtering in JPQ is a very subtle issue, because the soundness and the consistency of the semantics for filtering elements in the hierarchical structure and the optional structure are quite complex. For theoretical details of data filtering on these complex structures, please refer to our technique report\cite{report}.
%These issues are very different from those in conventional query languages which only concern filtering on plain tuples.

%\subsection{Function Declaration}
%\input{function.tex}
\section{Conclusion}

Document databases are becoming popular in practical data management. However, existing query mechanisms or query languages for JSON-like documents are not expressive enough to present complex document queries in practice. In this paper, we introduced a new query language named JPQ. JPQ is a pattern-based functional language which adopts various expressive patterns to extract structural data elements from JSON-like documents and to construct document fragments based on a deductive mechanism on data transformation. In comparison with the other query languages for JSON-like documents, JPQ exhibits many expressive and interesting features by the coherent pattern-based mechanisms in data extraction, data transformation and data filtering, as Fig.2 shows.

Our study on the JPQ language is still in its preliminary stage. Currently we have finished the design of the core language, and a prototype  based on the operational semantics is being implemented. However, although a sound semantics has been developed, how to efficiently process complex queries with compound conditions is still a tough problem to be solved.

\begin{figure}[h]
\begin{center}
\includegraphics[width=.9\textwidth]{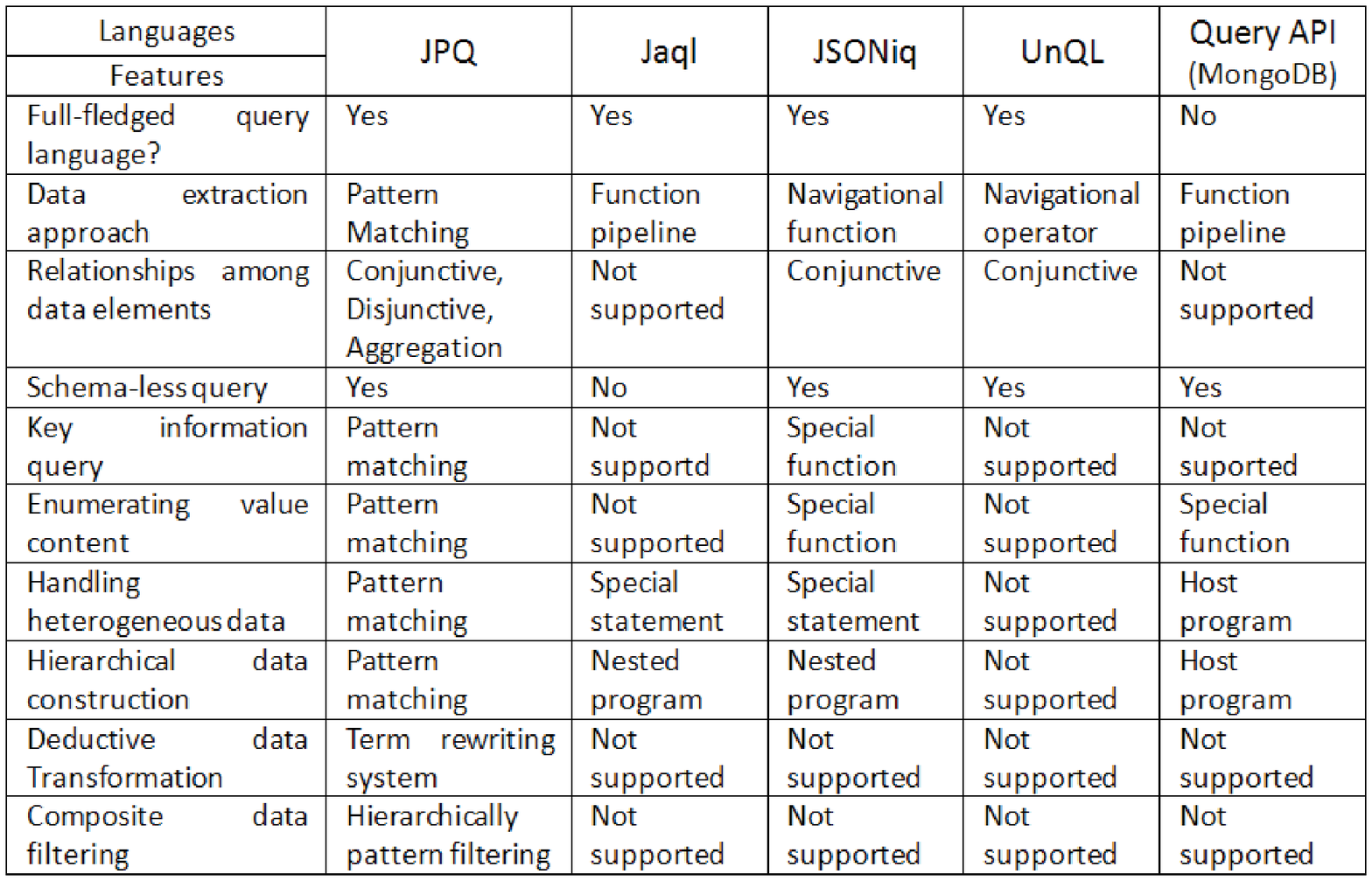}
\caption{Comparison of JSON-like document data query languages and interfaces}
\end{center}
\end{figure}

The study on some deeper topics of JPQ is also underway. We are extending the language with the update function so as to make it a full-fledged manipulation language. Meanwhile, processing JPQ queries in parallel databases,  especially on a Map/Reduce framework, is also an interesting topic that we concern.

\end{document}